\documentclass{article}  
\usepackage{jamaica04}
\usepackage{graphicx}

\frompage{000} \topage{000}                                              

\def\rightmark{Energy dependence of hadronic observables}
\def\leftmark{C. Blume}
 
\title{Energy dependence of hadronic observables in central Pb+Pb
reactions at the CERN SPS} 

\authors{
{Christoph Blume$^1$ for the NA49 Collaboration$^a$
}\\[2.812mm]
{\normalsize
\hspace*{-8pt}$^1$ Institut f\"ur Kernphysik der J.W. Goethe Universit\"at, \\ 
60486 Frankfurt am Main, Germany
} 
}
 
\abstract{}

\keyword{Spectra, Fluctuations, HBT} 
\PACS{25.75Nq}
 
\begin{document}
 
\maketitle
\setcounter{page}{1}

\section{Introduction}

In the recent years the NA49 experiment has collected data on Pb+Pb collisions
at beam energies between 20 to 158 $A$GeV with the objective to cover the critical region
of energy densities where the expected phase transition to a deconfined
phase might occur in the early stage of the reactions. 
In this contribution the energy dependence of
various hadronic observables is presented. These include $m_{t}$- and
rapidity-distributions, particle ratios and particle ratio fluctuations,
as well as HBT radii.
NA49 is a fixed target experiment at the CERN SPS and consists of
a large acceptance magnetic spectrometer equipped with four TPCs as tracking devices
and a forward calorimeter for centrality selection.
Details on the experimental setup can be found in \cite{na49nim}.

\section{Particle spectra}

\begin{figure}[htb]
\begin{center}
\includegraphics[height=60mm]{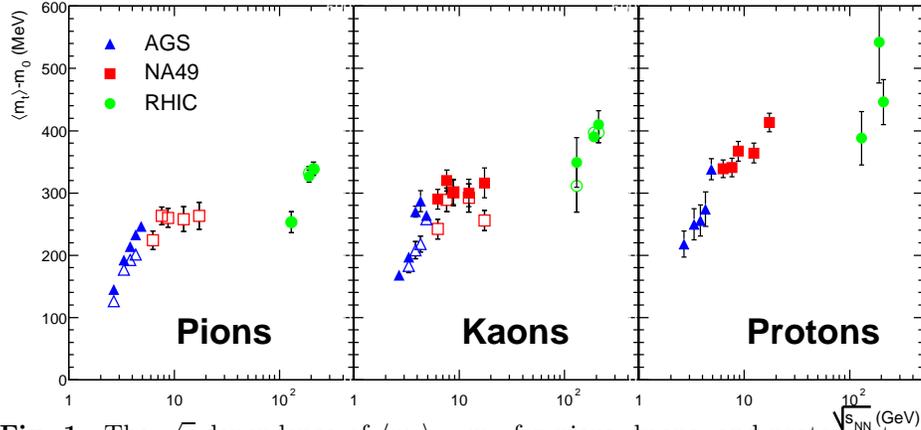}
\end{center}
\vspace*{-1cm}
\caption[]{The $\sqrt{s}$-dependence of $\langle m_{t} \rangle - m_{0}$ for
pions, kaons, and protons at mid-rapidity for 5-10\% most central
Au+Au/Pb+Pb reactions.}
\label{meanmt}
\end{figure}

Characterizing the energy dependence of the shape of
$m_{t}$-spectra generally requires a comparison to a model. A simple
exponential fit to the data can be misleading, since most particle
spectra exhibit a clear curvature. Pion $m_{t}$ spectra are concave, 
mainly due to resonance contributions, while proton spectra measured
in heavy ion reactions are convex, caused by the effect of radial flow.
To a certain extent, kaons are an exception that can relatively
well be approximated by an exponential. An analysis of the slope 
parameter for kaons, extracted by an exponential fit, revealed a 
clear change of its energy dependence around beam energies of 
20-30~$A$GeV \cite{marekqm}. Figure~\ref{meanmt} summarizes the
energy dependence of $\langle m_{t} \rangle - m_{0}$. This quantity
has the advantage of providing a model independent characterization of
the transverse mass distributions and can therefore easily be employed
for pions and protons as well. As can be seen, the change in the
evolution of the $m_{t}$-spectra with energy is clearly also present
for pions, similar to the kaons, and in a less pronounced fashion
for the protons as well.

\begin{figure}[htb]
\begin{center}
\includegraphics[height=90mm]{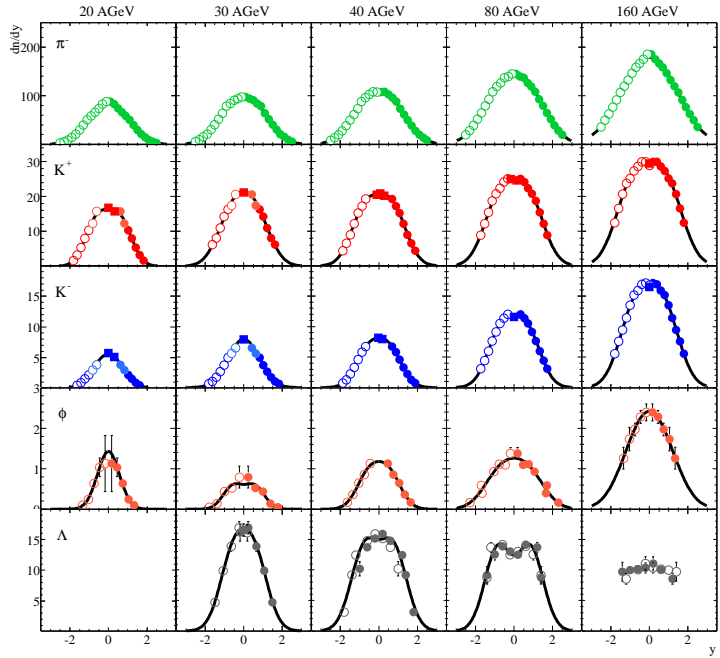}
\end{center}
\vspace*{-1.0cm}
\caption[]{The rapidity spectra of hadrons produced in central 
(7\% at 20-80~$A$GeV, 5\% ($\pi^{-}$, $K^{+}$, $K^{-}$) and
10\% ($\phi$, $\Lambda$) at 158~$A$GeV. The closed symbols indicate
measured points, open points are reflected with respect to mid-rapidity. 
The solid lines indicate parametrizations of the data used for
the extrapolation of the yield to full phase space.}
\label{rapidity}
\end{figure}

The large acceptance of the NA49 spectrometer allows to measure particle
spectra over a wide range of the longitudinal phase space. Figure~\ref{rapidity}
shows a compilation of the rapidity distributions of $\pi^{-}$, $K^{+}$, $K^{-}$,
$\phi$, and $\Lambda$ \cite{marekqm}. Generally, a clear increase of the widths
with beam energy can be observed, where the width of the $\pi^{-}$ distribution
is approximately equal to the $K^{+}$ and both are wider than the $K^{-}$ distribution.
The shape of the distribution for pions and kaons is well described by
a Gaussian. The $\Lambda$-distributions, however, exhibit a strong variation
of the shape: While at 30~$A$GeV they are still Gaussian-shaped, a clear
plateau develops with increasing beam energy.

\section{Particle multiplicities}

\begin{figure}[t]
\begin{minipage}[b]{45mm}
\begin{center}
\includegraphics[height=60mm]{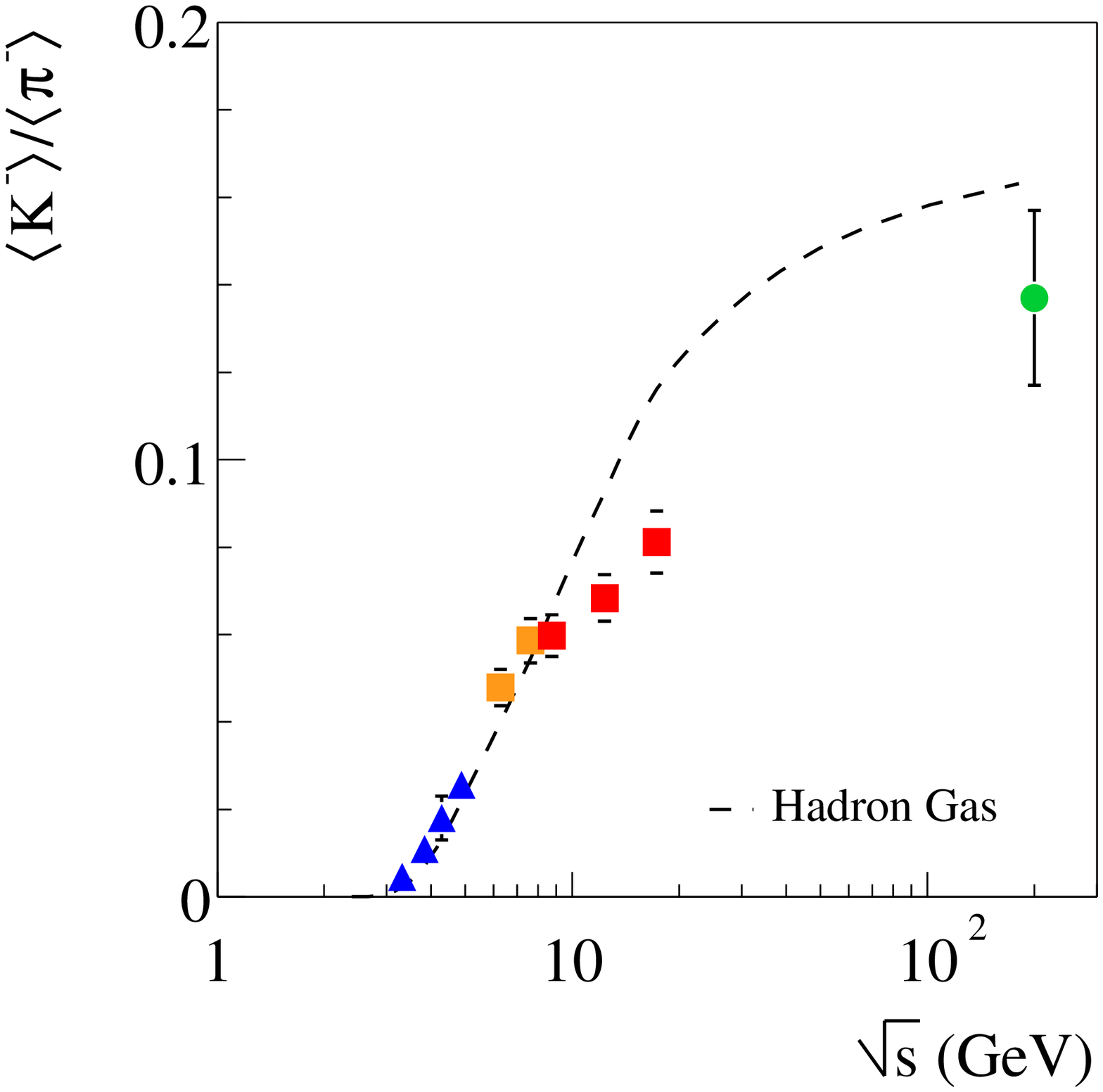}
\end{center}
\end{minipage}
%
\hspace{20mm}
\begin{minipage}[b]{45mm}
\begin{center}
\includegraphics[height=60mm]{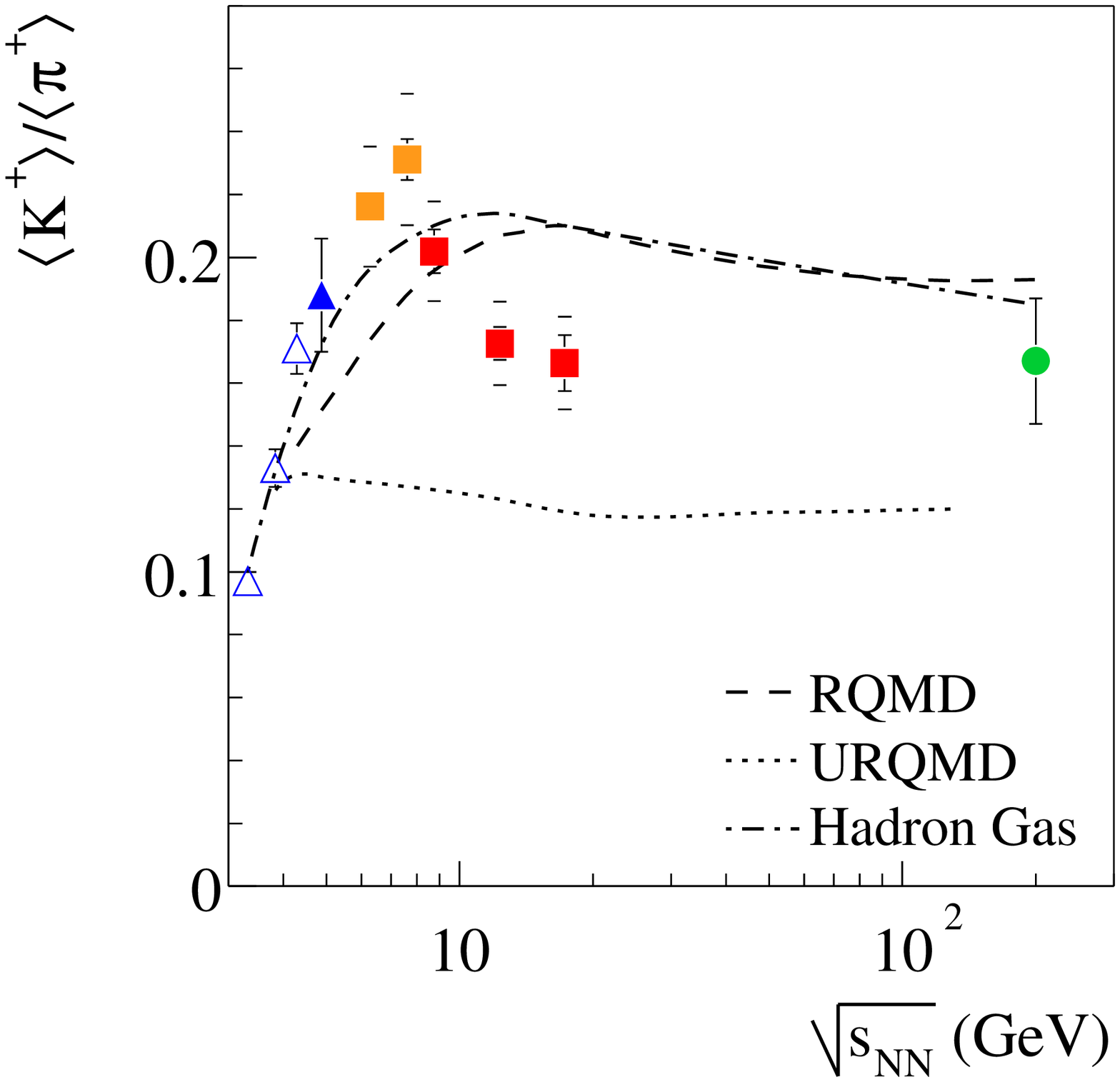}
\end{center}
\end{minipage}
\vspace*{-0.5cm}
\caption{The energy dependence of the $\langle K \rangle / \langle \pi \rangle$
ratios together with various model predictions (see text).}
\label{kpivssqrts}
\end{figure}

The rapidity distributions, discussed above, allow to 
determine the total yields of the different particle
species with only small extrapolations. Fig.\ref{kpivssqrts}
shows the ratio of the resulting 4$\pi$-yields of K and $\pi$ \cite{marekqm,na49kpi}. 
While this ratio for negatively charged particles rises more or less continuously
(left hand side of Fig.~\ref{kpivssqrts})
-- except a small indication for a kink at 30 $A$GeV --
a very distinct maximum is observed in the positively charged case
(right hand side of Fig.~\ref{kpivssqrts}).
The lines included in the figures are predictions of an extended hadron gas model 
\cite{pbm} and the
transport codes RQMD \cite{rqmd} and UrQMD \cite{urqmd}. Even though the hadron gas
model and RQMD predict a maximum of the 
$\langle K^{+} \rangle / \langle \pi^{+} \rangle$-ratio
in the SPS energy range, none of the models can fully describe the sharp feature of its
energy dependence. A feature, which is also not present in p+p collisions. 
It is also noteworthy that the hadron gas model does not fit the 
$\langle K^{-} \rangle / \langle \pi^{-} \rangle$-ratio at energies above 40~$A$GeV either.
On the other hand, a strong non-monotonic energy dependence 
of the total strangeness to pion ratio was
predicted by the Statistical Model of The Early Stage \cite{marek}, assuming a phase
transition from confined matter to a quark-gluon plasma in the SPS energy range.

\section{Particle ratio fluctuations}

\begin{figure}[t]
\vspace*{-2.2cm}
\begin{center}
\begin{minipage}[b]{50mm}
\begin{center}
\insertplot{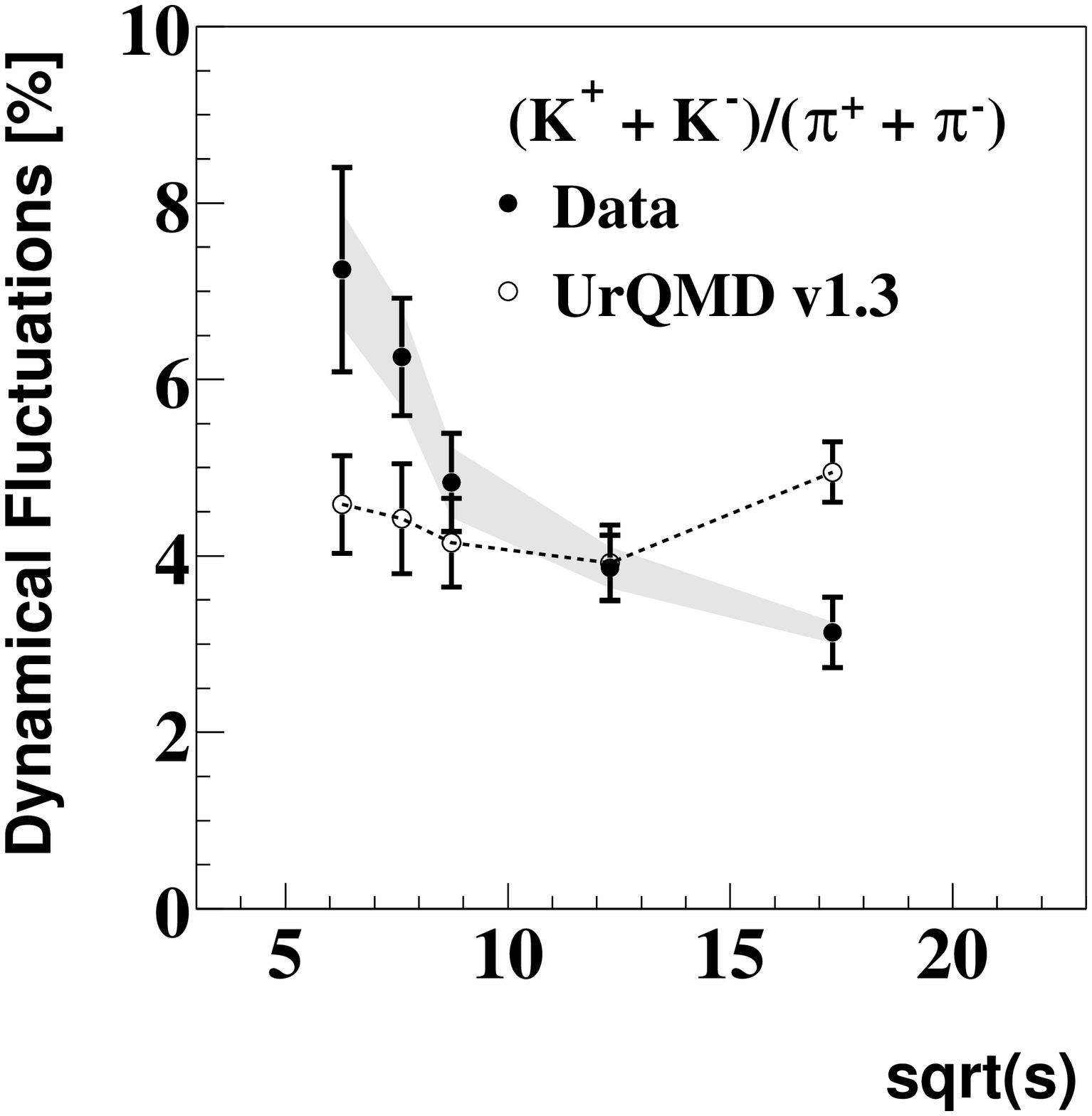}
\end{center}
\end{minipage}
%
\hspace{10mm}
\begin{minipage}[b]{50mm}
\begin{center}
\insertplot{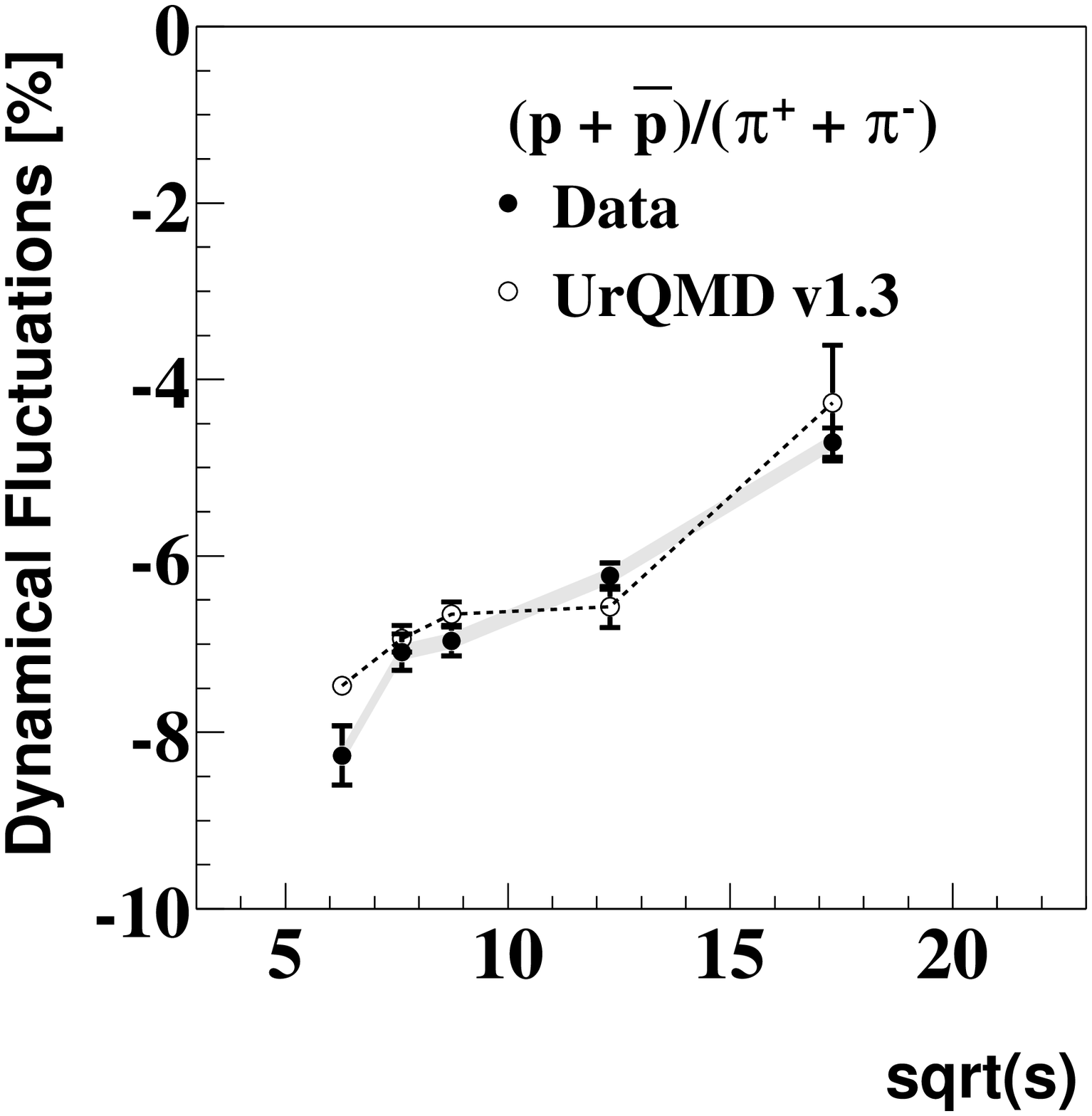}
\end{center}
\end{minipage}
\end{center}
\caption{Energy dependence of the event-by-event fluctuation signal of
the $(K^{+} + K^{-})/(\pi^{+} + \pi^{-})$ ratio (left hand side) and
the $(p + \bar{p})/(\pi^{+} + \pi^{-})$ ratio (right hand side). The 
systematic errors of the measurements are shown as gray bands.}
\label{fluct}
\end{figure}

NA49 has performed an event-by-event measurement of the particle ratios \cite{roland}.
The dynamical fluctuations $\sigma_{dyn}$ of this ratios have been extracted by 
subtracting the r.m.s. width $\sigma_{mix}$ of the mixed event distributions from the
r.m.s. width $\sigma_{data}$ of the real event distributions:
\begin{equation}
\sigma_{dyn} = sign(\sigma^{2}_{data} - \sigma^{2}_{mixed}) 
               \sqrt{|\sigma^{2}_{data} - \sigma^{2}_{mixed}|}
\end{equation} 
The mixed event particle ratios contain by construction the effects of finite number
fluctuations as well as effects of the detector resolution.
As shown in the left panel of Fig.~\ref{fluct}, the $K/\pi$ fluctuations are positive
and decrease with beam energy. The $p/\pi$ fluctuations, on the other hand, are negative
-- indicating a correlation present in the real data -- and increase with beam energies.
While the trend of the $K/\pi$ fluctuations is not reproduced by UrQMD \cite{urqmd2},
it provides a good description of the energy dependence of the $p/\pi$ fluctuations.
This might indicate that the negative value of the fluctuations in this ratio is 
due to resonance decays.

\section{Bose-Einstein correlations}

\begin{figure}[htb]
\insertplot{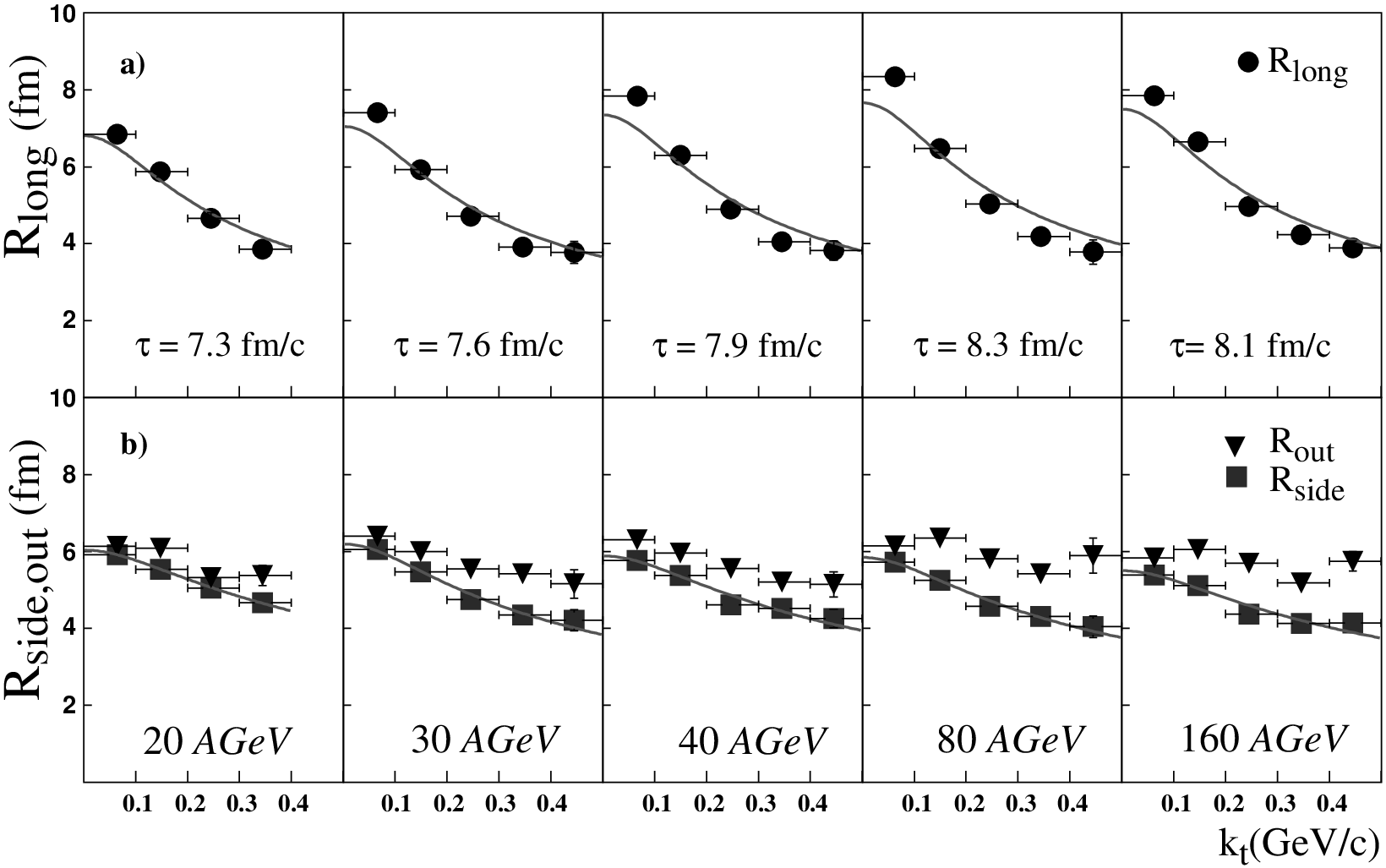}
\vspace*{-2cm}
\caption[]{The HBT-radii as a function of $k_{t}$ at mid-rapidity for 
central Pb+Pb collisions at 20 to 158 $A$GeV}
\label{hbtna49}
\end{figure}

Figure~\ref{hbtna49} summarizes the HBT-radii extracted in the LCMS, as
measured by the NA49 experiment \cite{stefan}. As expected in the
presence of longitudinal and transverse expansion, a significant
reduction of the radii with increasing $k_{t}$ is observed at all
beam energies. Assuming a boost-invariant scenario, the $k_{t}$-dependence
of $R_{long}$ should reflect the life time of the source \cite{rlong}:
\begin{equation}
R_{long} = \tau_{f} \left(\frac{T_{f}}{m_{t}}\right)^{1/2}\:; \:\:
m_{t} = (m_{\pi}^{2} + k_{t}^{2})^{1/2}
\end{equation}
The fits of this function, assuming a freeze-out temperature $T_{f} =$ 120 MeV,
are shown in the upper part of Fig.~\ref{hbtna49}. Only a weak increase of
the extracted life time with beam energy is observed.
Another important feature of this data is the fact that $R_{out} > R_{side}$
at all beam energies (lower part of Fig.~\ref{hbtna49}). 
The difference of these two parameter is connected to
the emission duration \cite{deltat}:
\begin{equation}
\Delta \tau^{2} = \frac{1}{\beta_{t}^{2}} (R_{out}^{2} - R_{side}^{2})\:; \:\:
\beta_{t} \approx \frac{k_{t}}{m_{t}}
\end{equation}
The data would indicate an emission duration of 3-4 fm/c.

\section{Conclusions}\label{concl}

The recent study of the excitation functions of hadronic observables
in the SPS energy range has revealed a number of interesting and unexpected
results. This includes a step-like energy dependence of the $\langle m_{t} \rangle - m_{0}$
of pions and kaons and a sharp maximum in the strangeness to pion ratio.
The dynamical $K/\pi$ ratio fluctuations are positive and decrease with beam energy in the
range between 20 - 158 $A$GeV, while the $p/\pi$ ratio fluctuations are negative and increase.
The HBT-radii, however, do not exhibit a significant energy dependence in this energy range.
 
 
\section*{Acknowledgements}

This work was supported by the US Department of Energy
Grant DE-FG03- \\
97ER41020/A000,
the Bundesministerium fur Bildung und Forschung, Germany, 
the Polish State Committee for Scientific Research 
(2 P03B 130 23, SPB/CERN/P-03/Dz 446/2002-2004, 2 P03B 04123), 
the Hungarian Scientific Research Foundation (T032648, T032293, T043514),
the Hungarian National Science Foundation, OTKA, (F034707),
the Polish-German Foundation, and the Korea Research Foundation 
Grant (KRF-2003-070-C00015).

\section*{Notes}
\begin{notes}
\item[a] The NA49 collaboration:\\
{\small
C.~Alt$^{9}$, T.~Anticic$^{21}$, B.~Baatar$^{8}$,D.~Barna$^{4}$,
J.~Bartke$^{6}$, 
L.~Betev$^{9,10}$, H.~Bia{\l}\-kowska$^{19}$, A.~Billmeier$^{9}$,
C.~Blume$^{9}$,  B.~Boimska$^{19}$
% ...
}\\
{\footnotesize
% ...
$^{4}$KFKI Research Institute for Particle and Nuclear Physics, Budapest, Hungary.\\
% ...
$^{6}$Institute of Nuclear Physics, Cracow, Poland.\\
% ...
$^{8}$Joint Institute for Nuclear Research, Dubna, Russia.\\
$^{9}$Fachbereich Physik der Universit\"{a}t, Frankfurt, Germany.\\
$^{10}$CERN, Geneva, Switzerland.\\
% ...
$^{19}$Institute for Nuclear Studies, Warsaw, Poland.\\
% ...
$^{21}$Rudjer Boskovic Institute, Zagreb, Croatia.\\
}
\end{notes}

\vfill\eject

\end{document}